\documentclass[12pt]{article}

\usepackage{epsfig}
\usepackage{graphics}
\input epsf

\title{Deep inelastic beauty production at HERA \\
  in the $k_T$-factorization approach}

\author{A.V.~Lipatov, N.P.~Zotov}

\begin{document}

\maketitle

\begin{center}

{\it D.V.~Skobeltsyn Institute of Nuclear Physics,\\ 
M.V. Lomonosov Moscow State University,
\\119992 Moscow, Russia\/}\\[3mm]

\end{center}

\vspace{1cm}

\begin{center}

{\bf Abstract }

\end{center}

We calculate the cross section of beauty production in $ep$ deep inelastic
scattering at HERA collider in the framework of the $k_T$-factorization 
approach. The unintegrated gluon distributions in a proton are obtained 
from the full CCFM, from unified BFKL-DGLAP evolution equations as
well as from the Kimber-Martin-Ryskin prescription.
We investigate different production rates and study the
$b$-quark contribution to the inclusive proton structure 
function $F_2(x, Q^2)$ at small $x$ and at moderate and high 
values of $Q^2$. Our theoretical results are compared
with the recent experimental data taken by the H1 and ZEUS collaborations.
We demonstrate the importance of leading $\ln 1/x$ 
contributions in description of the HERA data.

\vspace{1cm}

\section{Introduction} \indent 

The beauty production at high energies is a subject
of intensive study from both theoretical and experimental
points of view~[1--9]. First measurements~[1] of the $b$-quark cross
sections at HERA were significantly higher than the QCD predictions 
calculated at next-to-leading order (NLO) approximation. 
Similar observations were made in hadron-hadron
collisions at Tevatron~[2] and also in photon-photon interactions
at LEP2~[3]. In last case, the theoretical NLO QCD predictions are
more than three standard deviations below the experimental data.
At Tevatron, recent analisys indicates that the overall description 
of the data can be improved~[10] by adopting the non-perturbative 
fragmentation function of the $b$-quark into the $B$-meson: an appropriate
treatment of the $b$-quark fragmentation properties considerably
reduces the disagreement between measured beauty cross section
and the results of corresponding NLO QCD calculations. Also latest measurements~[4, 5, 9]
of the beauty photoproduction at HERA are in a reasonable 
agreement with the NLO QCD predictions or somewhat higher. Some disagreement 
is observed mainly at small decay muon and/or associated jet transverse 
momenta~[4, 5, 9]. But the large excess of the first measurements over NLO QCD, 
reported by the H1 collaboration~[1], is not confirmed.

Recently there have been become available experimental
data~[6--9] on the $b$-quark production in deep inelastic scattering at HERA which taken 
by the H1 and ZEUS collaborations. The first measurements~[6, 7] of 
the beauty contribution 
to the inclusive proton structure function $F_2(x, Q^2)$ have been 
presented for small values of the Bjorken scaling variable $x$, namely
$2 \cdot 10^{-4} < x < 5 \cdot 10^{-3}$, and for moderate and high
values of the photon virtuality $Q^2$, namely $12 < Q^2 < 700$~GeV$^2$.
Also process $e + p \to e' + b + \bar b + X \to e' + {\rm jet} + \mu + X'$ has 
been measured~[8, 9] in the small $x$ region with at least one jet and a decay muon 
in the final state and still was not described in the framework 
of QCD theory.
Such processes are dominated by the photon-gluon fusion subprocess
$\gamma^* + g \to b + \bar b$ and therefore sensitive to the gluon density 
in a proton $x g(x,\mu^2)$.
It was claimed~[8, 9] that the NLO QCD calculations have some difficulties in 
description of the recent HERA data. The predictions at 
low values of $Q^2$, Bjorken $x$, muon 
transverse momentum and high values of jet transverse energy and
muon pseudo-rapidity is about two standard deviation below the
data. 

In the present paper to analyze the recent H1 and ZEUS 
data~[6--9] we use the so-called $k_T$-factorization~[11, 12] (or semi-hard~[13, 14]) 
approach of QCD which has been applied earlier, in particular, 
in description of the charm and beauty production at HERA~[15--21], Tevatron~[22--28] 
and LEP2~[19, 29, 30] colliders.
The $k_T$-factorization approach is based on the Balitsky-Fadin-Kuraev-Lipatov 
(BFKL)~[31] or Ciafaloni-Catani-Fiorani-Marchesini (CCFM)~[32] gluon evolution
which are valid at small $x$ since here large logarithmic terms proportional 
to $\ln 1/x$ are summed up to all orders of perturbation theory (in the leading 
logarithmic approximation). It is in contrast with the popular 
Dokshitzer-Gribov-Lipatov-Altarelli-Parizi (DGLAP)~[33] strategy where only large 
logarithmic terms proportional to $\ln \mu^2$ are taken into account. 
The basic dynamical quantity of the 
$k_T$-factorization approach is the so-called unintegrated 
(${\mathbf k}_T$-dependent) gluon distribution 
${\cal A}(x,{\mathbf k}_T^2,\mu^2)$ which determines the probability to find a 
gluon carrying the longitudinal momentum fraction $x$ and the transverse momentum 
${\mathbf k}_T$ at the probing scale $\mu^2$. 
The unintegrated gluon distribution can be obtained from the
analytical or numerical solution of the BFKL or CCFM evolution equations.
Similar to DGLAP, to calculate the cross sections of any physical process the 
unintegrated gluon density ${\cal A}(x,{\mathbf k}_T^2,\mu^2)$ has to be 
convoluted~[11--14] with the relevant partonic cross section $\hat \sigma$. But as the 
virtualities of the propagating gluons are no longer ordered, the partonic cross 
section has to be taken off mass shell (${\mathbf k}_T$-dependent). It is in clear 
contrast with the DGLAP scheme (so-called collinear factorization). Since gluons 
in initial state are not on-shell and are characterized by virtual masses 
(proportional to their transverse momentum), it also assumes a modification 
of their polarization density matrix~[13, 14]. In particular, the polarization 
vector of a gluon is no longer purely transversal, but acquires an admixture of 
longitudinal and time-like components. Other important properties of the 
$k_T$-factorization formalism are the additional contribution to the cross 
sections due to the integration over the ${\mathbf k}_T^2$ region above $\mu^2$
and the broadening of the transverse momentum distributions due to extra 
transverse momentum of the colliding partons.

As it was noted already, some applications of the $k_T$-factorization approach 
supplemented with the BFKL and CCFM evolution to the heavy (charm and beauty) 
quark production at high energies are widely discussed in the literature~[15--30]
(see also review~[34, 35]). It was shown~[24--28] that 
the beauty cross section 
at Tevatron can be consistently described in the framework of this 
approach. However, a substantial discrepancy between theory
and experiment is still found~[19, 29, 30] for the $b$-quark production in 
$\gamma \gamma$ collisions at LEP2, not being cured by
the $k_T$-factorization\footnote{Some discussions of this problem 
may be found in~[19, 30].}. At HERA, the inclusive beauty photoproduction
has been investigated~[16, 17, 19, 21, 25]. In~[17, 19, 25] comparisons with the 
first H1 measurements~[1] have been done.
In~[17, 25] the Monte-Carlo generator \textsc{Cascade}~[36] has been used 
to predict the cross section of the $b$-quark and dijet associated 
photoproduction. However, calculations~[17, 19, 25] deal with the total cross sections 
only. In our previous paper~[21] the total and differential cross sections of
beauty photoproduction (both inclusive and associated with 
hadronic jets) have been considered and comparisons with the 
recent H1 and ZEUS measurements~[1, 4, 5, 9] have been made.
It was demonstrated~[21] that the $k_T$-factorization approach supplemented with the 
CCFM or BFKL-DGLAP evolved unintegrated gluon distributions~[28, 37]
reproduces well the numerous HERA data~[1, 4, 5, 9].

In the present paper we will study the beauty production in $ep$ deep 
inelastic scattering at HERA. We investigate a number of different 
production rates (in particular, the transverse momentum and pseudo-rapidity 
distributions of muons which originate from the semi-leptonic decays of 
$b$-quarks). Our study is based on leading-order (LO) 
off-mass shell matrix elements for the photon-gluon fusion 
subprocess $e + g^* \to e' + b + \bar b$.
Particularly we discuss the photoproduction limit ($Q^2 \to 0$) of
our derivation.
Also we investigate the beauty contribution to the inclusive proton
structure function $F_2(x, Q^2)$. In the numerical analysis we test 
the unintegrated gluon distributions which were obtained~[28, 37] from the full CCFM, 
unified BFKL-DGLAP evolution equations and from the conventional parton 
densities (using the Kimber-Martin-Ryskin prescription~[38]). We attempted a 
systematic comparison of model predictions with the recent experimental
data~[6--9] taken by the H1 and ZEUS collaborations.
One of purposes of this paper is to investigate the 
specific $k_T$-factorization effects in the $b$-quark leptoproduction at HERA.

The outline of  our paper is following. In Section~2 we 
recall the basic formulas of the $k_T$-factorization approach with a brief 
review of calculation steps. In Section~3 we present the numerical results
of our calculations and a discussion. Finally, in Section~4, we give
some conclusions. The compact analytic expressions for the
off-mass shell matrix elements of the photon-gluon fusion 
subprocess $e + g^* \to e' + b + \bar b$ are given in Appendix.
These formulas may be useful for the subsequent applications.

\section{Calculation details} \indent 

In this section we present our analytic results for the
cross section of $e + p \to e' + b + \bar b + X$ in DIS.
We work at leading-order $k_T$-factorization
approach of QCD. We start by defining the kinematics.

\subsection{Kinematics} \indent 

We denote the four-momenta of the incoming electron
and proton and the outgoing electron, beauty quark and anti-quark
by $p_e$, $p_p$, $p_e'$, $p_b$ and $p_{\bar b}$, respectively. 
The off-shell gluon and virtual photon have
four-momenta $k$ and $q$, and it is customary to define
$$
  k^2 = k_T^2 = - {\mathbf k}_T^2 < 0, \quad q^2 = (p_e - p_e')^2 = q_T^2 = 
    -Q^2 < 0, \eqno(1)
$$

\noindent 
where $k_{T}$ and $q_{T}$ are the transverse four-momenta of the corresponding 
particles. Choosing a suitable coordinate system in the $ep$ center-of-mass frame, 
we have
$$
  p_e = {\sqrt s}/2\,(1,0,0,-1),\quad p_p = {\sqrt s}/2\,(1,0,0,1), \eqno(2)
$$

\noindent 
where $\sqrt s$ is the total energy of the process under consideration
and we neglect the masses of the incoming electron and proton. The
standard deep inelastic variables $x$ and $y$ are defined as usual:
$$
  x = {Q^2\over 2 (p_p \cdot q)}, \quad y = {(p_p \cdot q)\over (p_e \cdot p_p)} 
    \simeq {Q^2\over x s}. \eqno(3)
$$

\noindent 
The variable $y$ measures the relative electron energy loss in the proton
rest frame. From the conservation law we can easily obtain the following 
condition:
$$
  \quad {\mathbf k}_{T} + {\mathbf q}_{T} = {\mathbf p}_{b\, T} + 
    {\mathbf p}_{\bar b \, T}. \eqno(4)
$$

\subsection{Cross section for deep inelastic beauty production} \indent 

According to the $k_T$-factorization theorem, the cross 
section of deep inelastic beauty production
$e + p \to e' + b + \bar b + X$ can be written as a convolution
$$
  \sigma(e + p\to e' + b + \bar b + X) = \int {dx\over x} 
    {\cal A}(x,{\mathbf k}_{T}^2,\mu^2) d{\mathbf k}_{T}^2 {d\phi \over 2\pi} 
    d \hat \sigma(e + g^* \to e' + b + \bar b), \eqno(5)
$$

\noindent
where ${\cal A}(x,{\mathbf k}_{T}^2,\mu^2)$ is the unintegrated
gluon distribution in a proton, $\hat \sigma(e + g^* \to e' + b + \bar b)$
is the cross section of partonic subprocess and $\phi$ is the 
azimuthal angle of initial virtual gluon.
Decomposing the cross section $\hat \sigma(e + g^* \to e' + b + \bar b)$ 
into a leptonic and a hadronic part, we can write it as
$$
  d \sigma(e + g^* \to e' + b + \bar b) = {1\over 64 x s} {e^2\over Q^4} L^{\mu \nu} 
    H_{\mu \nu}\, d\Phi^{(3)}(p_e + k, p_e', p_b, p_{\bar b}), \eqno(6)
$$

\noindent
where $e$ is the electron charge magnitude and $L^{\mu \nu}$ and $H^{\mu \nu}$ are
the leptonic and hadronic tensors. In general case the Lorentz-invariant element 
$d\Phi^{(n)}(p, p_1, \ldots, p_n)$ of $n$-body phase space is given by
$$
  d \Phi^{(n)}(p, p_1, \ldots, p_n) = (2 \pi^4)\,\delta^{(4)}\left(p - \sum_{i = 1}^n p_i\right) \prod_{i = 1}^n {d^3 p_i\over (2\pi)^3 2 p_i^0}. \eqno(7)
$$

\noindent
Integrating over the azimuthal angle of the outgoing electron, we can
simplify (6) to become
$$
  d \sigma(e + g^* \to e' + b + \bar b) = {\alpha\over 2\pi} {1\over 64 xs} L^{\mu \nu} H_{\mu \nu} {dy\over y} {d Q^2\over Q^2} d\Phi^{(2)}(q + k, p_b, p_{\bar b}), \eqno(8)
$$

\noindent
where $\alpha = e^2/(4 \pi)$ is Sommerfeld's fine structure constant.
For the leptonic tensor $L^{\mu \nu}$ we use the following expression~[39]:
$$
  L^{\mu \nu} = {1 + (1 - y)^2\over y}\epsilon^{\mu \nu}_T - 
    {4(1 - y)\over y}\epsilon^{\mu \nu}_L, \eqno(9)
$$

\noindent
where
$$
  \displaystyle \epsilon^{\mu \nu}_T = - g^{\mu \nu} + {q^\mu k^\nu + 
    q^\nu k^\mu\over (q \cdot k)} - {q^2\over (q \cdot k)^2} k^\mu k^\nu, \atop {
    \displaystyle \epsilon^{\mu \nu}_L = {1\over q^2}\left(q^\mu - 
    {q^2\over (q \cdot k)} k^\mu \right)\left(q^\nu - 
    {q^2\over (q \cdot k)} k^\nu \right)}. \eqno(10)
$$

\noindent
The $\epsilon^{\mu \nu}_T$ and $\epsilon^{\mu \nu}_L$ refer to transverse
and longitudinal virtual photon polarization, as indicated by their subscripts.
It is easily to see that $q_\mu \epsilon^{\mu \nu}_T = q_\mu \epsilon^{\mu \nu}_L = 0$,
$\epsilon^{\mu}_{\mu\,T} = - 2$ and $\epsilon^{\mu}_{\mu\,L} = - 1$.
Furthermore,
$$
  \epsilon^{\mu \nu} = \epsilon^{\mu \nu}_T + \epsilon^{\mu \nu}_L = - g^{\mu \nu} + 
    {q^\mu q^\nu \over q^2}, \eqno(11)
$$

\noindent
i.e. $\epsilon^{\mu \nu}$ is the polarization tensor of an unpolarized 
spin-one boson having mass $q^2$. From (5) --- (10) one can obtain
the following formula for the cross section of deep inelastic beauty 
production in the $k_T$-factorization approach:
$$
  \displaystyle \sigma(e + p \to e' + b + \bar b + X) = \int {1\over 256 \pi^3 (x y s)^2}
    {\cal A}(x,{\mathbf k}_{T}^2,\mu^2) \times \atop 
    \displaystyle \times \left[{(1 + (1 - y)^2)\over Q^2} T({\mathbf k}_T^2, Q^2) - 
    4(1 - y) L({\mathbf k}_T^2, Q^2)\right] d{\mathbf p}_{b\,T}^2 d{\mathbf k}_{T}^2 dQ^2
    dy_b dy_{\bar b} {d\phi \over 2\pi} {d\phi_b \over 2\pi} 
    {d\phi_{\bar b} \over 2\pi}, \eqno(12)
$$

\noindent
where $y_b$, $y_{\bar b}$ and $\phi_b$ and $\phi_{\bar b}$ are the 
rapidities and azimuthal angles of the produced beauty quark and anti-quark, 
respectively. The evaluation of functions $T({\mathbf k}_T^2, Q^2)$ and 
$L({\mathbf k}_T^2, Q^2)$ has been done analytically
using the \textsc{Mathematica 5} program.
The compact expressions for these functions are listed in Appendix.
It is important that the functions $T({\mathbf k}_T^2, Q^2)$ and 
$L({\mathbf k}_T^2, Q^2)$ depend on the virtual gluon 
non-zero transverse momentum ${\mathbf k}_{T}^2$. 
Note that if we average (12) over ${\mathbf k}_{T}$ and 
take the limit ${\mathbf k}_{T}^2 \to 0$, then we obtain well-known 
formula corresponding to the usual LO QCD calculations.

It is interesting to study the photoproduction limit of (12) by taking the
limit $Q^2 \to 0$. This provides us with a powerful check for our 
formulas by relating them to well-known results. So, the cross
section of the partonic process $\gamma + g^* \to b + \bar b$ reads
$$
  d \sigma(\gamma + g^* \to b + \bar b) = {1\over 64 \hat s} (-g^{\mu \nu}) 
    H_{\mu \nu}\biggm|_{Q^2 = 0} d\Phi^{(2)}(q + k, p_b, p_{\bar b}), \eqno(13)
$$

\noindent
where $\hat s = (q + k)^2$. Comparing (8) and (13), one can obtain
the well-known relation
$$
  \lim_{Q^2 \to 0} Q^2 {d\sigma(e + g^* \to e' + b + \bar b)\over dy dQ^2} = 
    {\alpha\over 2\pi} {1 + (1 - y)^2\over y} \sigma(\gamma + g^* \to b + 
    \bar b). \eqno(14)
$$

\noindent
The contribution of $b$-quarks to the deep inelastic proton
structure function $F_2(x, Q^2)$ can be calculated according
to convolution (5) also. The relevant coefficient function
is described by the quark box diagram and has been presented
in our previous paper~[40].

The multidimensional integration in (12) has been performed
by means of the Monte Carlo technique, using the routine 
\textsc{Vegas}~[41]. The full C$++$ code is available from the authors on 
request\footnote{lipatov@theory.sinp.msu.ru}.

\section{Numerical results} \indent 

We now are in a position to present our numerical results. 
First we describe our theoretical input and the kinematical conditions. 

\subsection{Theoretical uncertainties} \indent 

There are several parameters which determined the normalization factor of 
the cross section (12): the beauty 
mass $m_b$, the factorization and 
normalisation scales $\mu_F$ and $\mu_R$ and the unintegrated gluon distributions 
in a proton ${\cal A}(x,{\mathbf k}_T^2,\mu^2)$. 

Concerning the unintegrated gluon densities in a proton, in the numerical calculations 
we used five different sets of them, namely the J2003 (set~1 --- 3)~[28], KMS~[37]
and KMR~[38]. All these distributions are widely discussed in the literature 
(see, for example, review~[34, 35] for more information). Here we only shortly 
discuss their characteristic properties.
First, three sets of the J2003 gluon density have been obtained~[28] 
from the numerical solution of the full CCFM equation. The input parameters were fitted to 
describe the proton structure function $F_2(x,Q^2)$.
Note that the J2003~set~1 and J2003 set~3 densities contain only
singular terms in the CCFM splitting function $P_{gg}(z)$. 
The J2003~set~2 gluon density takes into account the additional non-singlular 
terms\footnote{See Ref.~[28] for more details.}. These distributions have been 
applied in the analysis of the forward jet production at HERA and charm and bottom 
production at Tevatron~[28] (in the framework of Monte-Carlo generator \textsc{Cascade}~[36]) 
and have been used also in our calculations~[20, 21].

Another set (the KMS)~[37] was obtained from a unified 
BFKL-DGLAP description of $F_2(x, Q^2)$ data and includes the so-called 
consistency constraint~[42]. The consistency constraint introduces a 
large correction to the LO BFKL equation. It was argued~[43] that 
about 70\% of the full NLO corrections to the BFKL exponent 
$\Delta$ are effectively included in this constraint. 
The KMS gluon density is successful in description of the 
beauty hadroproduction at Tevatron~[24, 26] and photoproduction
at HERA~[21].

The last, fifth unintegrated gluon distribution ${\cal A}(x,{\mathbf k}_T^2,\mu^2)$
used here (the so-called KMR distribution) 
is the one which was originally proposed in~[38]. The KMR approach is the formalism 
to construct unintegrated gluon distribution from the known conventional parton
(quark and gluon) densities. It accounts for the angular-ordering (which comes from the coherence 
effects in gluon emission) as well as the main part of the collinear higher-order QCD 
corrections. The key observation here is that the $\mu$ dependence of the unintegrated 
parton distribution enters at the last step of the evolution, and therefore single scale 
evolution equations (DGLAP or unified BFKL-DGLAP) can be used up to this step. 
Also it was shown~[38] that the unintegrated distributions obtained via unified BFKL-DGLAP 
evolution are rather similar to those based on the pure DGLAP equations.
It is because the imposition of the angular ordering constraint is more 
important~[38] than including the BFKL effects. Based on this point, 
in our further calculations we use much more simpler DGLAP equation up to 
the last evolution step\footnote{We have used the standard GRV (LO) parametrizations~[44]
of the collinear quark and gluon densities.}. Note that the KMR parton densities 
in a proton were used, in particular, to describe the prompt photon photoproduction at HERA~[45] 
and prompt photon hadroproduction Tevatron~[46, 47].

Also the significant theoretical uncertainties in our results connect with the
choice of the factorization and renormalization scales. First of them
is related to the evolution of the gluon distributions, the other is 
responsible for the strong coupling constant $\alpha_s(\mu^2_R)$.
The optimal values of these scales are such that the contribution of higher
orders in the perturbative expansion is minimal.
As it often done for beauty production, we choose the renormalization and 
factorization scales to be equal: 
$\mu_R = \mu_F = \mu = \sqrt{m_b^2 + \langle {\mathbf p}_{T}^2 \rangle}$, 
where $\langle {\mathbf p}_{T}^2 \rangle$ is set to the average ${\mathbf p}_{T}^2$ 
of the beauty quark and antiquark. But in the case of the KMS gluon distribution 
we used special choice $\mu^2 = {\mathbf k}_T^2$, as it was originally proposed in~[37]. 
Note that in the present paper we concentrate mostly on the non-collinear gluon 
evolution in the proton and do not study the scale dependence of our 
results. To completeness, we take the $b$-quark mass 
$m_b = 4.75$~GeV and use LO formula for the coupling constant $\alpha_s(\mu^2)$ 
with $n_f = 4$ active quark flavours at $\Lambda_{\rm QCD} = 200$~MeV, such 
that $\alpha_s(M_Z^2) = 0.1232$.

\subsection{Associated beauty and jet production} \indent 

The recent experimental data~[8, 9] for the associated beauty and hadronic jet 
leptoproduction at HERA comes from both the H1 and ZEUS collaborations.
The total and differential cross sections as a function of the
photon virtuality $Q^2$, Bjorken scaling variable $x$, muon transverse 
momentum $p_T^\mu$ and pseudo-rapidity $\eta^\mu$ and jet transverse momentum 
$p_T^{\rm jet}$ have been determined.
The ZEUS data~[8] refer to the kinematical region\footnote{Here and in the following 
all kinematic quantities are given in the laboratory frame where positive OZ axis 
direction is given by the proton beam.} defined by $Q^2 > 2$~GeV$^2$ with
at least one hadron-level jet (in the Breit frame) with 
$p_T^{\rm jet \, Breit} > 6$~GeV and $-2 < \eta^{\rm jet} < 2.5$ and with
muon which fulfill the following conditions: $-0.9 < \eta^\mu < 1.3$ and
$p_T^\mu > 2$~GeV or $-1.6 < \eta^\mu < -0.9$ and $p^\mu > 2$~GeV.
The fraction $y$ of the electron energy transferred to the photon
is restricted to the range $0.05 < y < 0.7$.
Note that the Breit frame is defined by the usual 
condition ${\mathbf q} + 2 x {\mathbf p_p} = 0$. 
In this frame, a space-like photon and proton collide head-on and any 
final-state particle with a high transverse momentum is produced by a hard 
QCD interaction.
The more recent H1 data~[9] refer to the kinematical region
defined by $2 < Q^2 < 100$~GeV$^2$, $0.1 < y < 0.7$, $p_T^\mu > 2.5$~GeV,
$-0.75 < \eta^\mu < 1.15$, $p_T^{\rm jet \, Breit} > 6$~GeV and 
$|\eta^{\rm jet}| < 2.5$. To produce muons from $b$-quarks in our theoretical 
calculations, we first convert $b$-quarks into $B$-hadrons using
the Peterson fragmentation function~[48] and then 
simulate their semileptonic decay according to the
standard electroweak theory. 
Of course, the muon transverse momenta spectra are 
sensitive to the fragmentation functions. However, this dependence is 
expected to be small as compared with the uncertainties coming from the unintegrated 
gluon densities in a proton. Our default set 
of the fragmentation parameter is $\epsilon_b = 0.0035$.

The basic photon-gluon fusion subprocess $\gamma^* + g^* \to b + \bar b$
give rise to two high-energy $b$-quarks, which can further evolve into hadron jets.
In our calculations we assumed that the produced quarks (with their known kinematical 
parameters) are taken to play the role of the final jets.
These two quarks are accompanied by a number of gluons radiated 
in the course of the gluon evolution. As it has been noted in~[15], on
the average the gluon transverse momentum decreases from the hard interaction
block towards the proton. We assume that the gluon 
emitted in the last evolution step and having the four-momenta $k'$ 
compensates the whole transverse momentum of the gluon participating in the hard 
subprocess, i.e. ${\mathbf k'}_{T} \simeq - {\mathbf k}_{T}$. All the other emitted 
gluons are collected together in the proton remnant, which is 
assumed\footnote{Note that such assumption is also used in the KMR formalism.} 
to carry only a negligible transverse momentum compared to ${\mathbf k'}_{T}$. 
This gluon gives rise to a final hadron jet with $p_T^{\rm jet} = |{\mathbf k'}_{T}|$ 
in addition to the jet produced in the hard subprocess. From these three hadron 
jets we choose the one carrying the largest transverse momentum (in the Breit frame), 
and then compute the beauty and associated jet production rates.

The results of our calculations are shown in Figs.~1 --- 10 in comparison to 
the H1 and ZEUS experimental data~[8, 9] fo the $b$-quark and associated 
jet production. Solid, 
dashed, dash-dotted, dotted and short dash-dotted curves correspond to the 
predictions obtained with the J2003~set~1 --- 3, KMR and KMS unintegrated 
gluon densities, respectively. One can see that the overall agreement
between our results (calculated using the J2003 and KMS gluon densities) 
and experimental data~[8, 9] is a rather good. However, the measured
cross section as a function of the muon transverse momentum $p_T^\mu$
shows a slightly steeper behaviour than the theoretical predictions: the results of 
our calculations tends to underestimate the data at low $p_T^\mu$ (see Figs.~1 and~2). 
But in general these predictions
still agree with the H1 and ZEUS data within the experimental 
uncertainties. Note also that the measured differential 
cross sections $d\sigma/d\eta^\mu$ in Figs.~3 and~4
exhibit a rise towards the forward muon pseudo-rapidity region, which 
is not reproduced~[8, 9] by the collinear NLO calculations. At the same time
the shape and the normalization of $\eta^\mu$ distributions are well 
described by our calculations. The collinear NLO QCD underestimate also the data
at low $Q^2$ and low $x$ values: it was claimed that in these kinematical regions 
the data are about two standard deviation higher~[8, 9].

As it was already mentioned above, the absolute normalization of the predicted cross 
sections in the framework of $k_T$-factorization approach is  
depends on the unintegrated gluon distribution used.
From Figs.~1 --- 10 one can see that all three sets of the J2003 gluon density 
as well as the KMS one give rise to results which are rather close to each 
other. So, the difference in normalization between the KMS and J2003 predictions 
is rather small, is about 15\% only. The similar effect we have found~[21]
in the case of beauty photoproduction. However, it is in the contrast 
with the $D^*$ meson and dijet 
associated photoproduction, which has been investigated 
in our previous paper~[20].
It was demonstrated~[20] a relative large enhancement 
of the cross sections calculated using the KMS gluon density. The 
possible explanation of this fact is that the large $b$-quark mass 
(which provide a hard scale) makes predictions of the perturbation 
theory of QCD more applicable.
Note also that the KMS gluon density provides a more hard transverse momentum 
distribution of the final muon (or jet) as compared with other 
unintegrated densities under consideration. Similar
effect we have observed~[21] in the case of beauty photoproduction.

Concerning the KMR predictions, one can see that 
this unintegrated gluon distribution gives results which lie below the data and 
which are very similar to the collinear NLO QCD. Such observation 
coincides with the ones~[20, 21]. This fact confirms the assumption which 
was made in~[45] that the KMR formalism results in some underestimation 
of the predicted cross sections. Such underestimation can be explained 
by the fact that leading logarithmic terms proportional to $\ln 1/x$ are 
not included into the KMR approach. 

\begin{table}
\begin{center}
\begin{tabular}{|l|c|}
\hline
  Source & $\sigma(e + p \to e' + {\rm jet} + \mu + X)$~[pb] \\
\hline
  ZEUS measurement~[8] & $40.9 \pm 5.7~{\rm (stat.)}^{+6.0}_{-4.4}~{\rm (syst.)}$\\
  NLO QCD (\textsc{hvqdis}~[50]) & $20.6^{+3.1}_{-2.2}$ \\
  \textsc{Rapgap}~[48] & 14.0 \\
  \textsc{Cascade}~[36] & 28.0 \\
  J2003 set 1 & 35.27 \\
  J2003 set 2 & 33.47 \\
  J2003 set 3 & 36.75 \\
  KMR & 22.11 \\
  KMS & 38.52 \\
\hline
\end{tabular}
\end{center}
\caption{The total cross section of beauty and associated jet leptoproduction 
obtained in the kinematic range 
$Q^2 > 2$~GeV$^2$, $0.05 < y < 0.7$, $p_T^{\rm jet \, Breit} > 6$~GeV,
$-2 < \eta^{\rm jet} < 2.5$ and 
$p_T^\mu > 2$~GeV, $-0.9 < \eta^\mu < 1.3$ or 
$p^\mu > 2$~GeV, $-1.6 < \eta^\mu < -0.9$.}
\end{table}

Now we turn to the total cross section of $b$-quark and associated jet
leptoproduction. In Table~1 and~2 we compare our theoretical results
with the H1 and ZEUS data~[8, 9] obtained in relevant kinematical 
regions (defined above).
The predictions of Monte-Carlo generators \textsc{Rapgap}~[49], 
\textsc{Cascade}~[36] as well as NLO QCD calculations (\textsc{hvqdis} program)~[50] 
are also shown for comparison. One can see that the collinear NLO QCD predictions
is about 2.5 standard deviation lower than the ZEUS data and 
is about 1.8 standard deviation lower then the H1 data. 
At the same time, our predictions obtained using 
the J2003 and KMS gluon densities are significantly higher 
and agree well with the both H1 and ZEUS data within the experimental
uncertainties. The KMR unintegrated gluon distribution again gives the results 
which are below the data and which are very close to NLO QCD ones. 
Note that the Monte-Carlo
generators \textsc{Rapgap} and \textsc{Cascade} also predict
a lower cross section than that measured in the data.

In general, we can conclude that the cross sections of deep inelastic
beauty and associated jet production calculated in the $k_T$-factorization
formalism (supplemented with the CCFM or unified BFKL-DGLAP evolution)
are larger by $30 - 40$\% than ones calculated at 
NLO level of collinear QCD. 
This enhancement comes, in particular, from the non-zero
transverse momentum of the incoming off-shell gluons and 
from taken into account the leading $\ln 1/x$ terms.
Our results for the total and differential 
cross sections are in a better agreement (both in normalization and
shape) with the H1 and ZEUS data than 
the NLO QCD predictions.

\begin{table}
\begin{center}
\begin{tabular}{|l|c|}
\hline
  Source & $\sigma(e + p \to e' + {\rm jet} + \mu + X)$~[pb] \\
\hline
  H1 measurement~[9] & $16.3 \pm 2.0~{\rm (stat.)} \pm 2.3~{\rm (syst.)}$\\
  NLO QCD (\textsc{hvqdis}~[50]) & $9.0^{+2.6}_{-1.6}$ \\
  \textsc{Rapgap}~[49] & 6.3 \\
  \textsc{Cascade}~[36] & 9.8 \\
  J2003 set 1 & 19.96 \\
  J2003 set 2 & 18.98 \\
  J2003 set 3 & 20.80 \\
  KMR & 12.45 \\
  KMS & 22.61 \\
\hline
\end{tabular}
\end{center}
\caption{The total cross section of beauty and associated jet leptoproduction 
obtained in the kinematic range 
$2 < Q^2 < 100$~GeV$^2$, $0.1 < y < 0.7$, $p_T^\mu > 2.5$~GeV,
$-0.75 < \eta^\mu < 1.15$, $p_T^{\rm jet \, Breit} > 6$~GeV and 
$|\eta^{\rm jet}| < 2.5$.}
\end{table}

\subsection{Beauty contribution to the proton SF $F_2(x, Q^2)$} \indent 

Now we will concentrate on the $b$-quark contribution
to the inclusive proton structure function $F_2(x, Q^2)$.
We will use the master formulas which were obtained in
our previous paper~[40]. As it was mentioned above, the first experimental data~[6, 7] 
on the structure function $F_2^b(x, Q^2)$ comes
from the H1 collaboration. These data refer to the
kinematical region defined by $2 \cdot 10^{-4} < x < 5 \cdot 10^{-3}$
and $12 < Q^2 < 650$~GeV$^2$.

Note that we change now the default set of parameters which we
have used in the previous section. So, we set the renormalization
and factorization scales $\mu_R$ and $\mu_F$ to be equal
to photon virtuality $Q^2$, as it was done earlier in analysis~[51]
of the charm contribution to the structure function $F_2(x, Q^2)$
in the framework of $k_T$-factorization QCD approach. The
similar choice have been used also in the analysis of 
longitudinal structure function $F_L(x, Q^2)$~[52]. Of course,
in the case of the KMS gluon distribution we set $\mu^2_R = \mu^2_F = {\mathbf k}_T^2$,
as it was originally proposed in~[37]. Other parameters have not been changed.

In Fig.~11 we show the structure function $F_2^b(x, Q^2)$ as a 
function of $x$ for different values of $Q^2$ in comparison
to the recent H1 data~[6, 7]. One can see that the J2003 distributions 
reproduce well the experimental data for all values of $Q^2$.
The KMS gluon density demonstrates a perfect agreement with the data
at moderate $Q^2$ but slightly overestimate them at $Q^2 = 650$~GeV$^2$.
It is interesting to note that the KMR density does not 
contradict the experimental data, too. However,
this distribution predicts a more rapid rise
of the calculated function $F_2^b(x, Q^2)$ 
with decreasing of $x$ (in comparison to the J2003 and KMS densities). 
We can conclude that in the small $x$ 
region ($x < 10^{-2}$) the shape of function $F_2^b(x, Q^2)$ predicted
by the unintegrated gluon distributions under consideration
is very different. In particular, the differences observed between 
the curves are due to the different behaviour of the corresponding 
unintegrated gluon distributions as a function of $x$ and ${\mathbf k}_T^2$~[45].
This fact shows the importance of a detail understanding 
of the non-collinear parton evolution in a proton and
the necessarity of better experimental constraints as well 
as further theoretical studies in this field.

\section{Conclusions} \indent 

We have calculated the deep inelastic beauty and associated jet production in 
electron-proton collisions at HERA in the $k_T$-factorization QCD approach.
The total and several differential cross section (as a function of the
photon virtuality $Q^2$, Bjorken scaling variable $x$, decay muon transverse 
momentum $p_T^\mu$ and pseudo-rapidity $\eta^\mu$ and hadronic 
jet transverse momentum $p_T^{\rm jet}$) have been studied.
Additionally we have investigated the $b$-quark contribution
to the inclusive proton structure function $F_2(x, Q^2)$ at
small $x$ and at moderate and high $Q^2$.
In numerical analysis we have used the unintegrated gluon densities 
which are obtained from the full CCFM (J2003~set~1 --- 3), 
from unified BFKL-DGLAP evolution equations (KMS) as well as 
from the Kimber-Martin-Ryskin prescription.
Our investigations were based on the LO off-mass shell 
matrix elements for photon-gluon fusion subprocesses.

We have shown that the $k_T$-factorization approach supplemented with the CCFM or 
BFKL-DGLAP evolved unintegrated gluon distributions (the J2003 or KMS densities) 
reproduces well the numerous HERA data
on beauty and associated jet production. At the same time we have obtained that the
Kimber-Martin-Ryskin formalism results in some underestimation of the
cross sections. This shows the importance of a detail
understanding of the non-collinear parton evolution process.

\section{Acknowledgements} \indent 

The authors are very grateful to S.P.~Baranov and A.V.~Kotikov 
for their encouraging interest and very helpful discussions.
This research was supported in part by the 
FASI of Russian Federation (grant NS-1685.2003.2).

\section{Appendix} \indent 

Here we present the compact analytic expressions for the 
functions $T({\mathbf k}_T^2, Q^2)$ and $L({\mathbf k}_T^2, Q^2)$ 
which appear in (12). In the following, $\hat s$, $\hat t$ and $\hat u$ 
are usual Mandelstam variables for corresponding $\gamma^* + g^* \to b + \bar b$
subprocesses ($\hat s + \hat t + \hat u = 2m^2 - Q^2 - {\mathbf k}_T^2$)
and $m$ and $e_b$ is the mass and fractional electric charge of $b$-quark.
The exact expressions for the functions $T({\mathbf k}_T^2, Q^2)$ and 
$L({\mathbf k}_T^2, Q^2)$ can be presented as
$$
  \displaystyle T({\mathbf k}_T^2, Q^2) = (4\pi)^3 \alpha^2 \alpha_s(\mu^2) e_b^2\, { F_T(\hat s, \hat t, \hat u, {\mathbf k}_T^2, Q^2)\over 8 (\hat t - m^2)^2 (\hat u - m^2)^2 (\hat s + Q^2 + {\mathbf k}_T^2)^4}, \eqno(A.2)
$$
$$
  \displaystyle L({\mathbf k}_T^2, Q^2) = (4\pi)^3 \alpha^2 \alpha_s(\mu^2) e_b^2\, { F_L(\hat s, \hat t, \hat u, {\mathbf k}_T^2, Q^2)\over 8 (\hat t - m^2)^2 (\hat u - m^2)^2 (\hat s + Q^2 + {\mathbf k}_T^2)^4}, \eqno(A.3)
$$

\noindent
where
$$
  F_T(\hat s, \hat t, \hat u, {\mathbf k}_T^2, Q^2) = - 8 (4 {\mathbf k}_T^8 Q^4 (\hat t - \hat u)^2 + 2 {\mathbf k}_T^6 Q^2 (-8 m^8 + {\hat t}^4 + 4 Q^4 (\hat t - \hat u)^2 - 
$$
$$
  4 {\hat t}^3 \hat u - 2 {\hat t}^2 {\hat u}^2 - 4 \hat t {\hat u}^3 + {\hat u}^4 + 16 m^6 (\hat t + \hat u) + 4 Q^2 (\hat t - \hat u)^2 (\hat t + \hat u) - 8 m^4 ({\hat t}^2 + 
$$
$$
  4 \hat t \hat u + {\hat u}^2) - 8 m^2 (Q^2 (\hat t - \hat u)^2 - 2 \hat t \hat u (\hat t + \hat u))) + 
$$
$$
  (\hat s + Q^2 + {\mathbf k}_T^2)^2 (24 m^{12} + 8 m^{10} (Q^2 - 3 (\hat t + \hat u)) - 2 m^8 (2 Q^4 + 3 {\hat t}^2 + 
$$
$$ 
  22 \hat t \hat u + 3 {\hat u}^2 + 10 Q^2 (\hat t + \hat u)) - \hat t \hat u ({\hat t}^2 + {\hat u}^2) (2 Q^4 + 2 Q^2 (\hat t + \hat u) + (\hat t + \hat u)^2) + 
$$
$$
  m^2 ({\hat t}^5 + 13 {\hat t}^4 \hat u + 26 {\hat t}^3 {\hat u}^2 + 26 {\hat t}^2 {\hat u}^3 + 13 \hat t {\hat u}^4 + {\hat u}^5 + 2 Q^4 (\hat t + \hat u)^3 + 
$$
$$
  4 Q^2 \hat t \hat u (3 {\hat t}^2 + 4 \hat t \hat u + 3 {\hat u}^2)) + 4 m^6 (2 Q^4 (\hat t + \hat u) + Q^2 (3 {\hat t}^2 + 14 \hat t \hat u + 3 {\hat u}^2) + 
$$
$$
  4 ({\hat t}^3 + 6 {\hat t}^2 \hat u + 6 \hat t {\hat u}^2 + {\hat u}^3)) - m^4 (7 {\hat t}^4 + 56 {\hat t}^3 \hat u + 90 {\hat t}^2 {\hat u}^2 + 56 \hat t {\hat u}^3 + 7 {\hat u}^4 + 
$$
$$
  6 Q^4 (\hat t + \hat u)^2 + 2 Q^2 ({\hat t}^3 + 19 {\hat t}^2 \hat u + 19 \hat t {\hat u}^2 + {\hat u}^3))) - 2 {\mathbf k}_T^4 (8 m^{12} - 2 Q^8 (\hat t - \hat u)^2 - 
$$
$$
  4 Q^6 (\hat t - \hat u)^2 (\hat t + \hat u) + \hat t \hat u (\hat t + \hat u)^2 ({\hat t}^2 + {\hat u}^2) - 4 Q^4 ({\hat t}^4 - 2 {\hat t}^3 \hat u - 2 {\hat t}^2 {\hat u}^2 - 2 \hat t {\hat u}^3 + {\hat u}^4) - 
$$
$$
  Q^2 ({\hat t}^5 - 5 {\hat t}^4 \hat u - 4 {\hat t}^3 {\hat u}^2 - 4 {\hat t}^2 {\hat u}^3 - 5 \hat t {\hat u}^4 + {\hat u}^5) + 8 m^{10} (2 Q^2 - 3 (\hat t + \hat u)) + 2 m^8 (8 Q^4 - 
$$
$$
  12 Q^2 (\hat t + \hat u) + 15 (\hat t + \hat u)^2) - 4 m^6 (8 Q^4 (\hat t + \hat u) + 5 (\hat t + \hat u)^3 + 
$$
$$
  Q^2 (-5 {\hat t}^2 + 2 \hat t \hat u - 5 {\hat u}^2)) + m^4 (7 {\hat t}^4 + 32 {\hat t}^3 \hat u + 42 {\hat t}^2 {\hat u}^2 + 32 \hat t {\hat u}^3 + 7 {\hat u}^4 + 
$$
$$
  8 Q^4 ({\hat t}^2 + 10 \hat t \hat u + {\hat u}^2) - 2 Q^2 (5 {\hat t}^3 - 13 {\hat t}^2 \hat u - 13 \hat t {\hat u}^2 + 5 {\hat u}^3)) + m^2 (-{\hat t}^5 + 
$$
$$
  8 Q^6 (\hat t - \hat u)^2 - 9 {\hat t}^4 \hat u - 14 {\hat t}^3 {\hat u}^2 - 14 {\hat t}^2 {\hat u}^3 - 9 \hat t {\hat u}^4 - {\hat u}^5 + 8 Q^4 ({\hat t}^3 - 5 {\hat t}^2 \hat u - 
$$
$$
  5 \hat t {\hat u}^2 + {\hat u}^3) + 4 Q^2 ({\hat t}^4 - 5 {\hat t}^3 \hat u - 4 {\hat t}^2 {\hat u}^2 - 5 \hat t {\hat u}^3 + {\hat u}^4))) + {\mathbf k}_T^2 (32 m^{14} - 
$$
$$ 
  112 m^{12} (\hat t + \hat u) - 8 m^{10} (4 Q^4 - 17 {\hat t}^2 - 50 \hat t \hat u - 17 {\hat u}^2) - 
$$
$$
  2 \hat t \hat u (\hat t + \hat u)^3 ({\hat t}^2 + {\hat u}^2) + Q^2 ({\hat t}^2 - {\hat u}^2)^2 ({\hat t}^2 - 4 \hat t \hat u + {\hat u}^2) + 
$$
$$
  2 Q^6 ({\hat t}^4 - 4 {\hat t}^3 \hat u - 2 {\hat t}^2 {\hat u}^2 - 4 \hat t {\hat u}^3 + {\hat u}^4) + 2 Q^4 ({\hat t}^5 - 5 {\hat t}^4 \hat u - 
$$
$$
  4 {\hat t}^3 {\hat u}^2 - 4 {\hat t}^2 {\hat u}^3 - 5 \hat t {\hat u}^4 + {\hat u}^5) - 4 m^8 (4 Q^6 + 19 {\hat t}^3 - 
$$
$$
  14 Q^2 (\hat t - \hat u)^2 + 121 {\hat t}^2 \hat u + 121 \hat t {\hat u}^2 + 19 {\hat u}^3 - 12 Q^4 (\hat t + \hat u)) + 4 m^6 (5 {\hat t}^4 + 
$$
$$
  72 {\hat t}^3 \hat u + 126 {\hat t}^2 {\hat u}^2 + 72 \hat t {\hat u}^3 + 5 {\hat u}^4 + 8 Q^6 (\hat t + \hat u) - 20 Q^2 (\hat t - \hat u)^2 (\hat t + \hat u) - 
$$
$$
  2 Q^4 (5 {\hat t}^2 - 2 \hat t \hat u + 5 {\hat u}^2)) - 2 m^4 ({\hat t}^5 + 49 {\hat t}^4 \hat u + 118 {\hat t}^3 {\hat u}^2 + 118 {\hat t}^2 {\hat u}^3 + 49 \hat t {\hat u}^4 + 
$$
$$ 
  {\hat u}^5 + 8 Q^6 ({\hat t}^2 + 4 \hat t \hat u + {\hat u}^2) - 3 Q^2 (\hat t - \hat u)^2 (7 {\hat t}^2 + 10 \hat t \hat u + 7 {\hat u}^2) - 
$$
$$
  2 Q^4 (5 {\hat t}^3 - 13 {\hat t}^2 \hat u - 13 \hat t {\hat u}^2 + 5 {\hat u}^3)) + 2 m^2 (16 Q^6 \hat t \hat u (\hat t + \hat u) + 
$$
$$
  2 \hat t \hat u (\hat t + \hat u)^2 (5 {\hat t}^2 + 4 \hat t \hat u + 5 {\hat u}^2) - Q^2 (\hat t - \hat u)^2 (5 {\hat t}^3 + 3 {\hat t}^2 \hat u + 
$$
$$
  3 \hat t {\hat u}^2 + 5 {\hat u}^3) - 4 Q^4 ({\hat t}^4 - 5 {\hat t}^3 \hat u - 4 {\hat t}^2 {\hat u}^2 - 5 \hat t {\hat u}^3 + {\hat u}^4)))), \eqno(A.4)
$$
$$
  F_L(\hat s, \hat t, \hat u, {\mathbf k}_T^2, Q^2) = 16 (2 {\mathbf k}_T^8 Q^2 (\hat t - \hat u)^2 + {\mathbf k}_T^6 (\hat t - \hat u)^2 (2 m^4 + 4 Q^4 + {\hat t}^2 + {\hat u}^2 + 
$$
$$
  4 Q^2 (\hat t + \hat u) - 2 m^2 (4 Q^2 +\hat  t + \hat u)) + 2 (m^2 - \hat t) (m^2 - \hat u) (\hat s + Q^2 + {\mathbf k}_T^2)^2 (2 m^6 + 
$$
$$
  m^4 (Q^2 - \hat t - \hat u) + \hat t \hat u (Q^2 + \hat t + \hat u) - m^2 (2 \hat t \hat u + Q^2 (\hat t + \hat u))) + {\mathbf k}_T^4 (-8 m^8 Q^2 + 
$$
$$
  2 Q^6 (\hat t - \hat u)^2 + 4 Q^4 (\hat t - \hat u)^2 (\hat t + \hat u) + (\hat t - \hat u)^4 (\hat t + \hat u) + 
$$
$$
  Q^2 (3 {\hat t}^4 - 6 {\hat t}^3 \hat u - 2 {\hat t}^2 {\hat u}^2 - 6 \hat t {\hat u}^3 + 3 {\hat u}^4) + 8 m^6 (-(\hat t - \hat u)^2 + 2 Q^2 (\hat t + \hat u)) - 
$$
$$
  2 m^4 (-4 (\hat t - \hat u)^2 (\hat t + \hat u) + Q^2 ({\hat t}^2 + 22 \hat t \hat u + {\hat u}^2)) - 2 m^2 (4 Q^4 (\hat t - \hat u)^2 + 
$$
$$
  2 (\hat t - \hat u)^2 ({\hat t}^2 + {\hat u}^2) + Q^2 (3 {\hat t}^3 - 11 {\hat t}^2 \hat u - 11 \hat t {\hat u}^2 + 3 {\hat u}^3))) + 2 {\mathbf k}_T^2 (4 m^{12} - 
$$
$$
  4 m^{10} (2 Q^2 + 3 (\hat t + \hat u)) + m^8 (-4 Q^4 + 17 {\hat t}^2 + 26 \hat t \hat u + 17 {\hat u}^2 + 12 Q^2 (\hat t + \hat u)) + 
$$
$$
  2 m^6 (4 Q^4 (\hat t + \hat u) - 5 (\hat t + \hat u)^3 - Q^2 ({\hat t}^2 + 6 \hat t \hat u + {\hat u}^2)) - \hat t \hat u (2 Q^4 ({\hat t}^2 + {\hat u}^2) + 
$$
$$
  (\hat t + \hat u)^2 ({\hat t}^2 - 3 \hat t \hat u + {\hat u}^2) + Q^2 (3 {\hat t}^3 + {\hat t}^2 \hat u + \hat t {\hat u}^2 + 3 {\hat u}^3)) + 2 m^2 (Q^4 (\hat t + \hat u)^3 + 
$$
$$
  \hat t \hat u ({\hat t}^3 - 7 {\hat t}^2 \hat u - 7 \hat t {\hat u}^2 + {\hat u}^3) + Q^2 ({\hat t}^4 + 5 {\hat t}^3 \hat u + 5 \hat t {\hat u}^3 + {\hat u}^4)) - m^4 (6 Q^4 (\hat t + \hat u)^2 + 
$$
$$
  Q^2 (5 {\hat t}^3 + 3 {\hat t}^2 \hat u + 3 \hat t {\hat u}^2 + 5 {\hat u}^3) - 2 ({\hat t}^4 + 5 {\hat t}^3 \hat u + 18 {\hat t}^2 {\hat u}^2 + 5 \hat t {\hat u}^3 + {\hat u}^4)))). \eqno(A.5)
$$

\newpage

\begin{figure}
\epsfig{figure=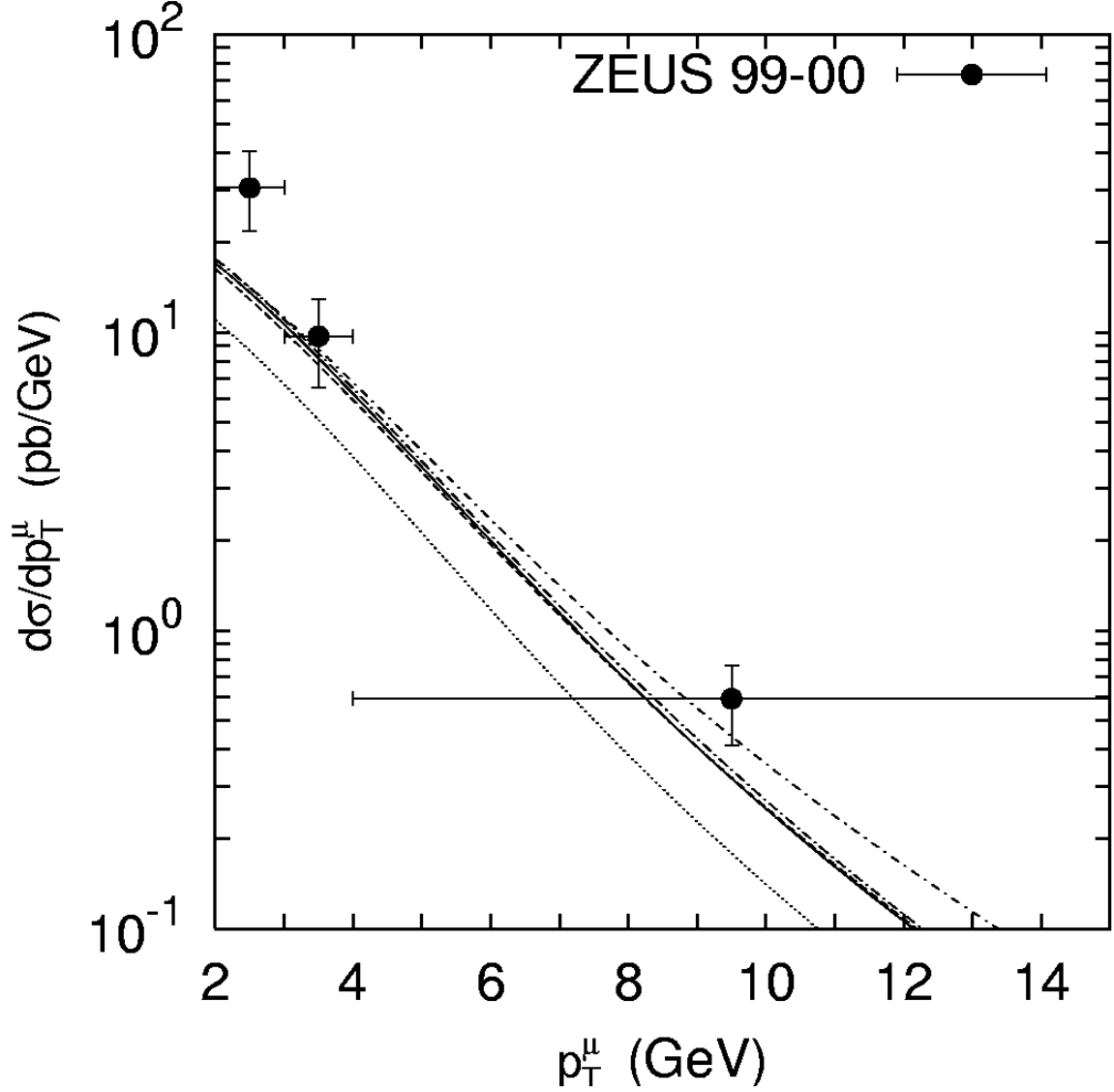, width = 22cm}
\caption{The muon transverse momentum distribution $d\sigma/d p_T^{\mu}$ of the 
deep inelastic beauty production at HERA in the kinematic range 
$Q^2 > 2$~GeV$^2$, $0.05 < y < 0.7$, $p_T^{\rm jet \, Breit} > 6$~GeV, 
$-2 < \eta^{\rm jet} < 2.5$ and $p_T^{\mu} > 2$~GeV,
$-0.9 < \eta^{\mu} < 1.3$ or $p^{\mu} > 2$~GeV,
$-1.6 < \eta^{\mu} < -0.9$. The solid, 
dashed, dash-dotted, dotted and short dash-dotted curves correspond to the 
predictions obtained with the J2003~set~1 --- 3, KMR and KMS unintegrated 
gluon densities, respectively.
The experimental data are from ZEUS~[8].}
\label{fig1}
\end{figure}

\newpage

\begin{figure}
\epsfig{figure=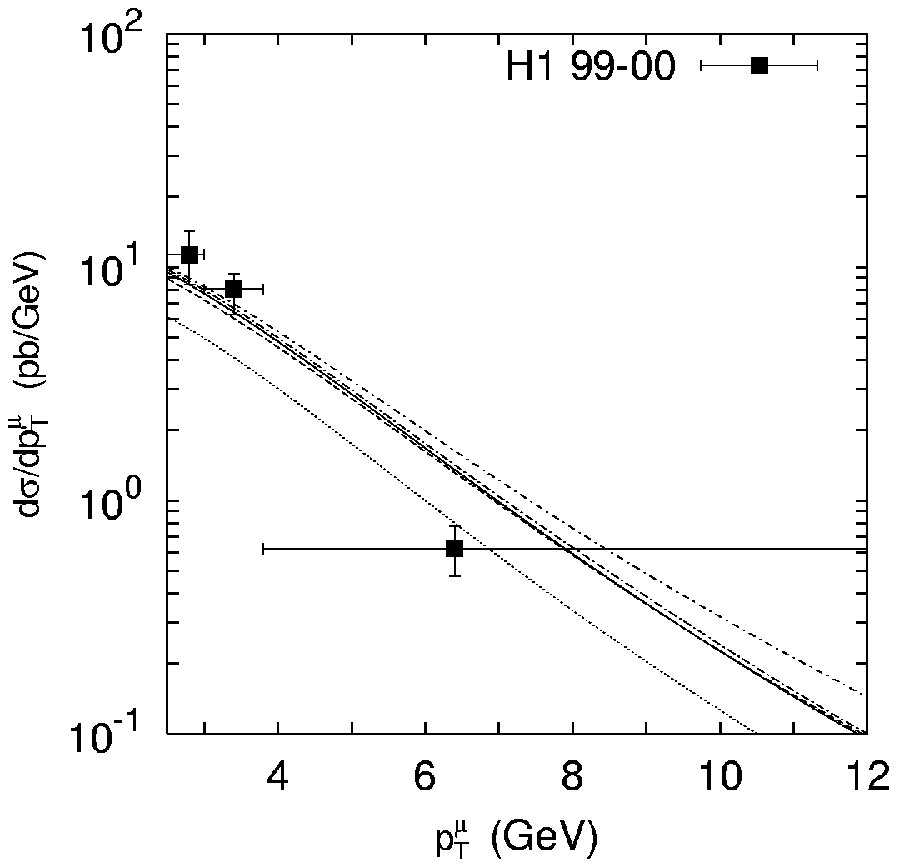, width = 22cm}
\caption{The muon transverse momentum distribution $d\sigma/d p_T^{\mu}$ of the 
deep inelastic beauty production at HERA in the kinematic range 
$2 < Q^2 < 100$~GeV$^2$, $0.1 < y < 0.7$, $p_T^{\rm jet \, Breit} > 6$~GeV, 
$|\eta^{\rm jet}| < 2$, $p_T^{\mu} > 2$~GeV and
$-0.75 < \eta^{\mu} < 1.15$. Notations of all curves are the 
same as in Fig.~1. The experimental data are from H1~[9].}
\label{fig2}
\end{figure}

\newpage

\begin{figure}
\epsfig{figure=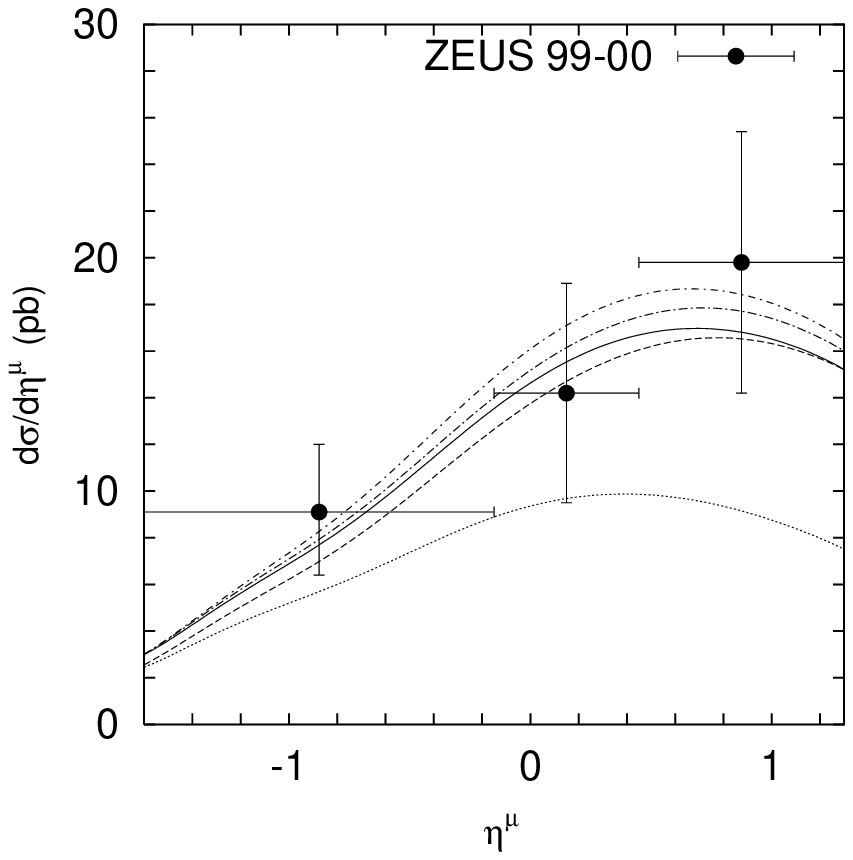, width = 22cm}
\caption{The muon pseudo-rapidity distribution $d\sigma/d \eta^{\mu}$ of the 
deep inelastic beauty production at HERA in the kinematic range 
$Q^2 > 2$~GeV$^2$, $0.05 < y < 0.7$, $p_T^{\rm jet \, Breit} > 6$~GeV, 
$-2 < \eta^{\rm jet} < 2.5$ and $p_T^{\mu} > 2$~GeV,
$-0.9 < \eta^{\mu} < 1.3$ or $p^{\mu} > 2$~GeV,
$-1.6 < \eta^{\mu} < -0.9$. Notations of all curves are the 
same as in Fig.~1.
The experimental data are from ZEUS~[8].}
\label{fig3}
\end{figure}

\newpage

\begin{figure}
\epsfig{figure=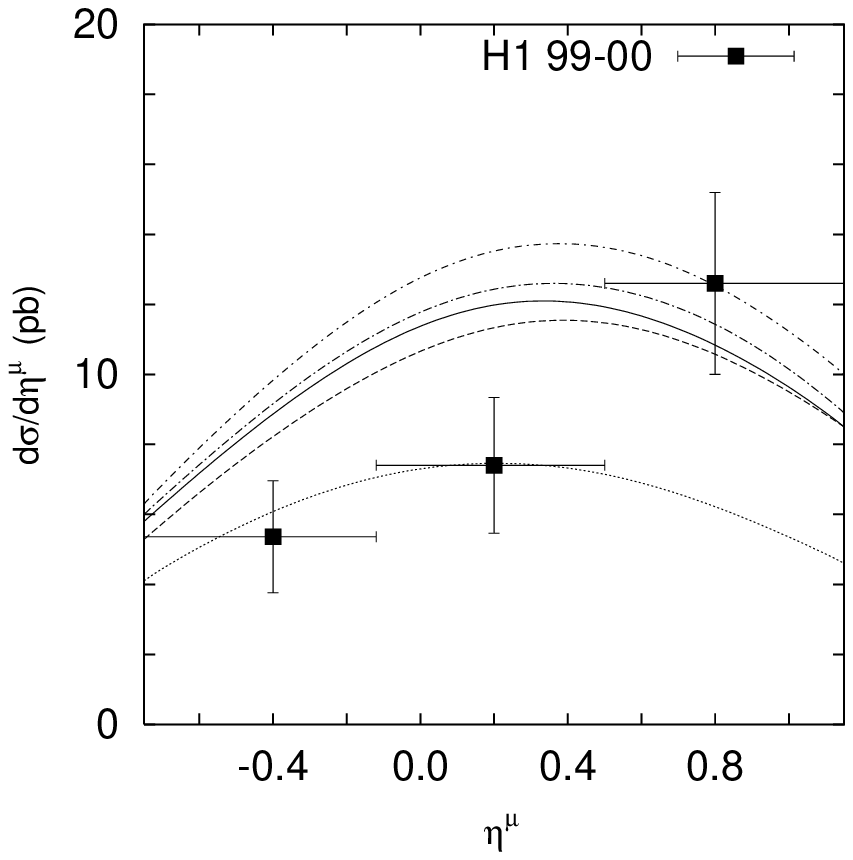, width = 22cm}
\caption{The muon pseudo-rapidity distribution $d\sigma/d \eta^{\mu}$ of the 
deep inelastic beauty production at HERA in the kinematic range 
$2 < Q^2 < 100$~GeV$^2$, $0.1 < y < 0.7$, $p_T^{\rm jet \, Breit} > 6$~GeV, 
$|\eta^{\rm jet}| < 2$, $p_T^{\mu} > 2$~GeV and
$-0.75 < \eta^{\mu} < 1.15$. Notations of all curves are the 
same as in Fig.~1. The experimental data are from H1~[9].}
\label{fig4}
\end{figure}

\newpage

\begin{figure}
\epsfig{figure=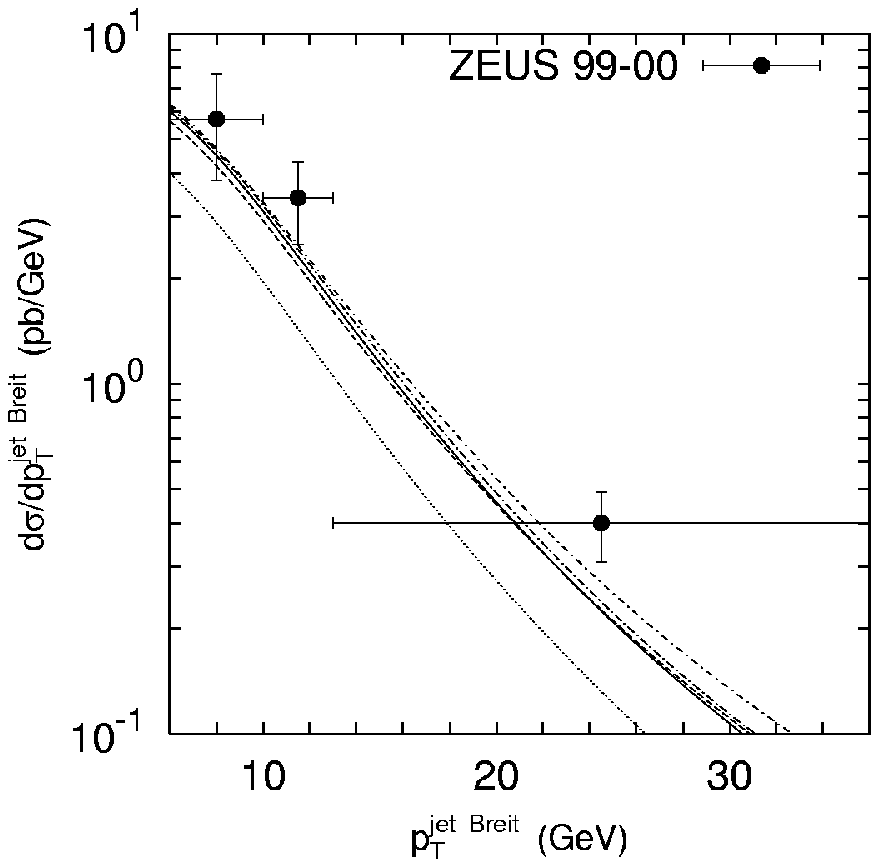, width = 22cm}
\caption{The jet transverse momentum distribution $d\sigma/d p_T^{\rm jet \, Breit}$ of the 
deep inelastic beauty production at HERA in the kinematic range 
$Q^2 > 2$~GeV$^2$, $0.05 < y < 0.7$, $p_T^{\rm jet \, Breit} > 6$~GeV, 
$-2 < \eta^{\rm jet} < 2.5$ and $p_T^{\mu} > 2$~GeV,
$-0.9 < \eta^{\mu} < 1.3$ or $p^{\mu} > 2$~GeV,
$-1.6 < \eta^{\mu} < -0.9$. Notations of all curves are the 
same as in Fig.~1.
The experimental data are from ZEUS~[8].}
\label{fig5}
\end{figure}

\newpage

\begin{figure}
\epsfig{figure=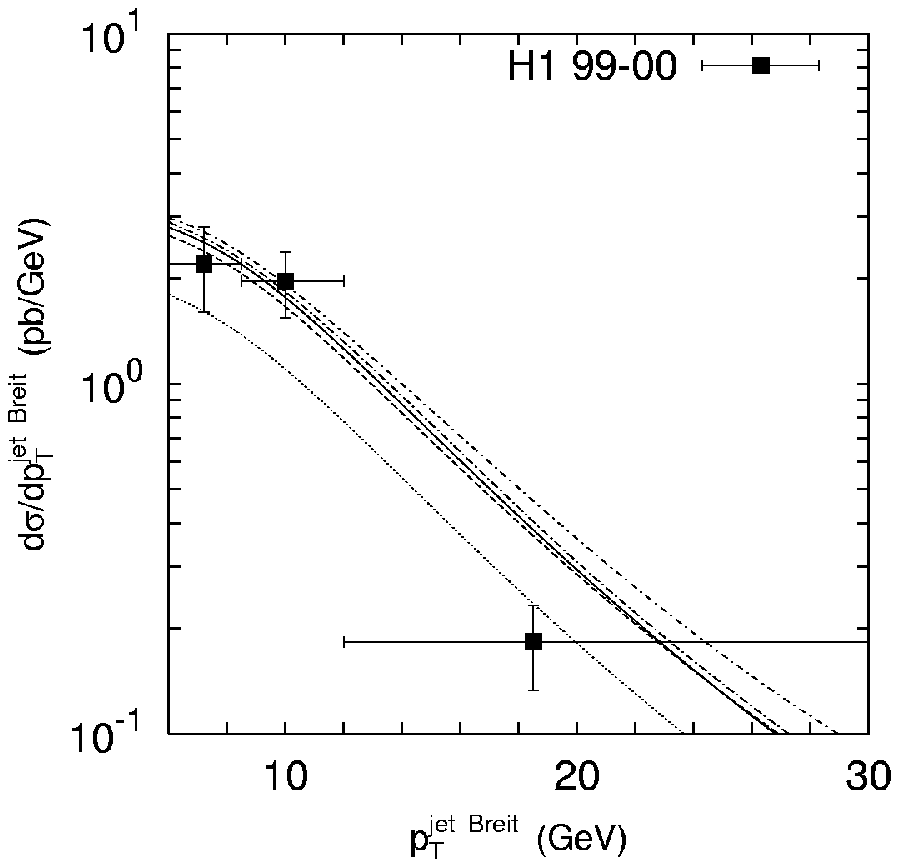, width = 22cm}
\caption{The jet transverse momentum distribution $d\sigma/d p_T^{\rm jet \, Breit}$ of the 
deep inelastic beauty production at HERA in the kinematic range 
$2 < Q^2 < 100$~GeV$^2$, $0.1 < y < 0.7$, $p_T^{\rm jet \, Breit} > 6$~GeV, 
$|\eta^{\rm jet}| < 2$, $p_T^{\mu} > 2$~GeV and
$-0.75 < \eta^{\mu} < 1.15$. Notations of all curves are the 
same as in Fig.~1. The experimental data are from H1~[9].}
\label{fig6}
\end{figure}

\newpage

\begin{figure}
\epsfig{figure=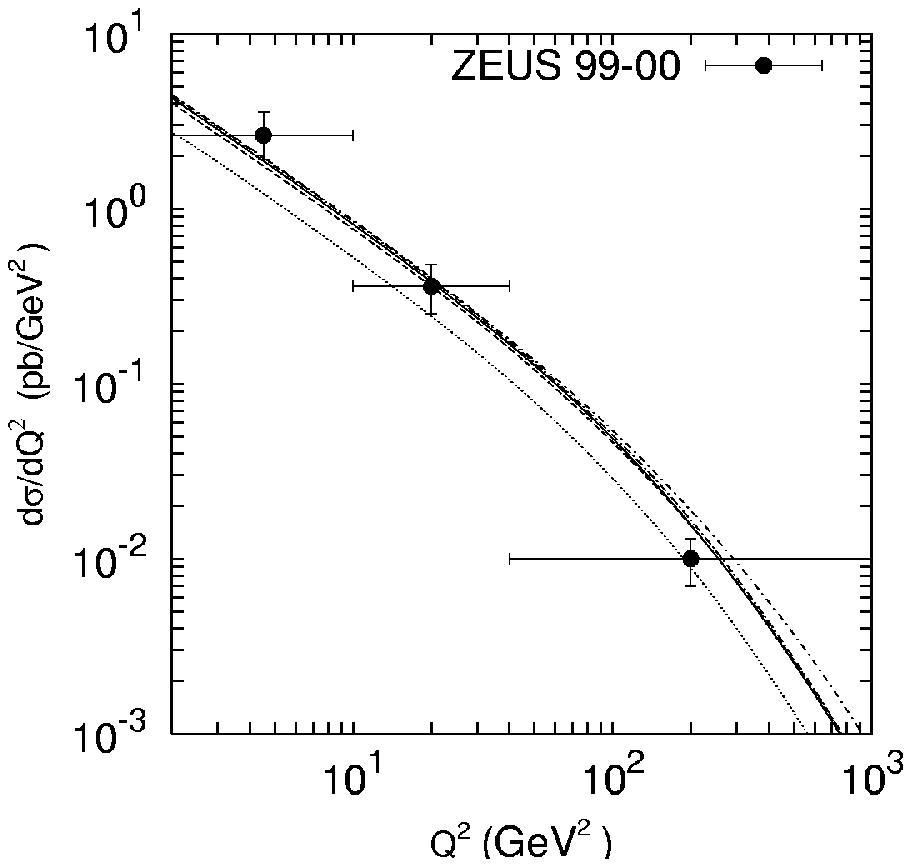, width = 22cm}
\caption{The $Q^2$ distribution of the
deep inelastic beauty production at HERA in the kinematic range 
$Q^2 > 2$~GeV$^2$, $0.05 < y < 0.7$, $p_T^{\rm jet \, Breit} > 6$~GeV, 
$-2 < \eta^{\rm jet} < 2.5$ and $p_T^{\mu} > 2$~GeV,
$-0.9 < \eta^{\mu} < 1.3$ or $p^{\mu} > 2$~GeV,
$-1.6 < \eta^{\mu} < -0.9$. Notations of all curves are the 
same as in Fig.~1.
The experimental data are from ZEUS~[8].}
\label{fig7}
\end{figure}

\newpage

\begin{figure}
\epsfig{figure=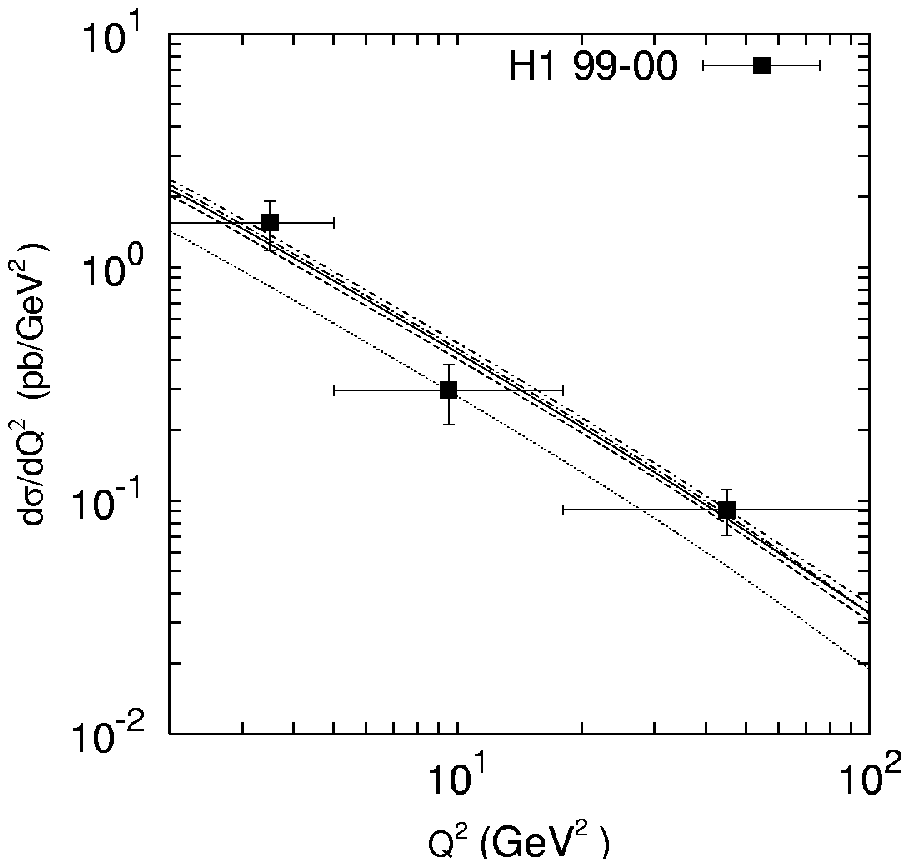, width = 22cm}
\caption{The $Q^2$ distribution of the 
deep inelastic beauty production at HERA in the kinematic range 
$2 < Q^2 < 100$~GeV$^2$, $0.1 < y < 0.7$, $p_T^{\rm jet \, Breit} > 6$~GeV, 
$|\eta^{\rm jet}| < 2$, $p_T^{\mu} > 2$~GeV and
$-0.75 < \eta^{\mu} < 1.15$. Notations of all curves are the 
same as in Fig.~1. The experimental data are from H1~[9].}
\label{fig8}
\end{figure}

\newpage

\begin{figure}
\epsfig{figure=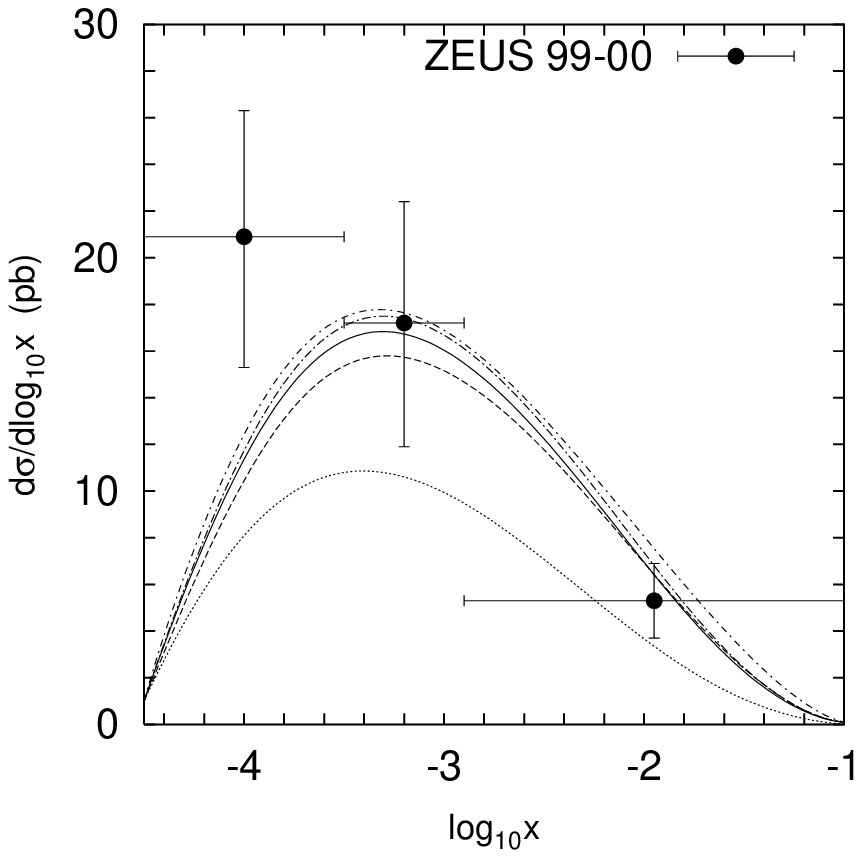, width = 22cm}
\caption{The $\log_{10} x$ distribution of the
deep inelastic beauty production at HERA in the kinematic range 
$Q^2 > 2$~GeV$^2$, $0.05 < y < 0.7$, $p_T^{\rm jet \, Breit} > 6$~GeV, 
$-2 < \eta^{\rm jet} < 2.5$ and $p_T^{\mu} > 2$~GeV,
$-0.9 < \eta^{\mu} < 1.3$ or $p^{\mu} > 2$~GeV,
$-1.6 < \eta^{\mu} < -0.9$. Notations of all curves are the 
same as in Fig.~1.
The experimental data are from ZEUS~[8].}
\label{fig9}
\end{figure}

\newpage

\begin{figure}
\epsfig{figure=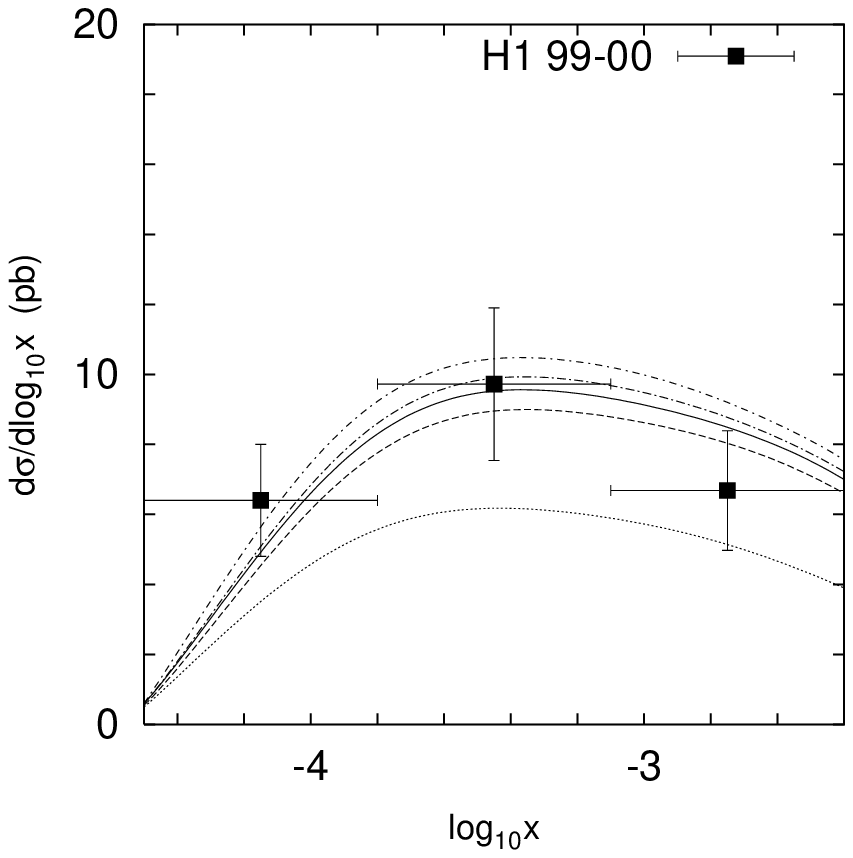, width = 22cm}
\caption{The $\log_{10} x$ distribution of the 
deep inelastic beauty production at HERA in the kinematic range 
$2 < Q^2 < 100$~GeV$^2$, $0.1 < y < 0.7$, $p_T^{\rm jet \, Breit} > 6$~GeV, 
$|\eta^{\rm jet}| < 2$, $p_T^{\mu} > 2$~GeV and
$-0.75 < \eta^{\mu} < 1.15$. Notations of all curves are the 
same as in Fig.~1. The experimental data are from H1~[9].}
\label{fig10}
\end{figure}

\newpage

\begin{figure}
\epsfig{figure=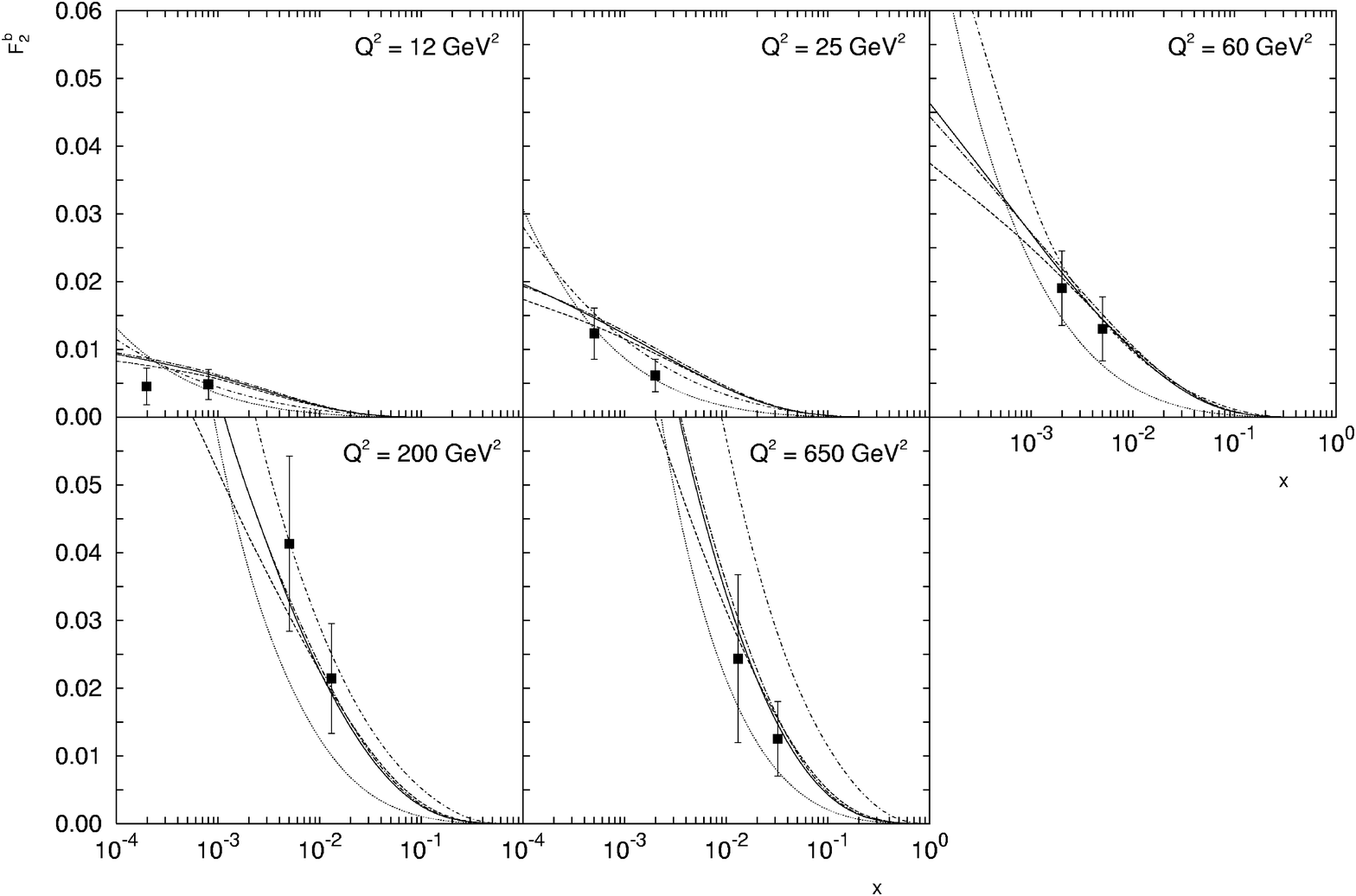, width = 17cm}
\caption{The structure function $F_2^b(x, Q^2)$ as a function of $x$ for
different values of $Q^2$. Notations of all curves are the 
same as in Fig.~1. The experimental data are from H1~[6, 7].}
\label{fig11}
\end{figure}


\begin{thebibliography}{46}

\bibitem{1} C.~Adloff {\sl et al.} (H1 Collaboration), Phys. Lett. B {\bf 467}, 156 (1999); Erratum: {\it ibid} B {\bf 518}, 331 (2001).
\bibitem{2} F.~Abe {\sl et al.} (CDF Collaboration), Phys. Rev. D {\bf 55}, 2546 (1997);\\
  D.~Acosta {\sl et al.} (CDF Collaboration), Phys. Rev. D {\bf 65}, 052002 (2002);\\
  S.~Abachi {\sl et al.} (D0 Collaboration), Phys. Lett. B {\bf 487}, 264 (2000).
\bibitem{3} M.~Acciari {\sl et al.} (L3 Collaboration), Phys. Lett. B {\bf 503}, 10 (2001);\\
  P.~Achard {\sl et al.} (L3 Collaboration), Phys. Lett. B {\bf 619}, 71 (2005);\\
  G.~Abbiendi {\sl et al.} (OPAL Collaboration), Eur. Phys. J. C {\bf 16}, 579 (2000).
\bibitem{4} J.~Breitweg {\sl et al.} (ZEUS Collaboration), Eur. Phys. J. C {\bf 18}, 625 (2001).
\bibitem{5} S.~Chekanov {\sl et al.} (ZEUS Collaboration), Phys. Rev. D {\bf 70}, 012008 (2004).
\bibitem{6} A.~Aktas {\sl et al.} (H1 Collaboration), hep-ex/0411046.
\bibitem{7} A.~Aktas {\sl et al.} (H1 Collaboration), hep-ex/0507081.
\bibitem{8} S.~Chekanov {\sl et al.} (ZEUS Collaboration), Phys. Lett. B {\bf 599}, 173 (2004).
\bibitem{9} A.~Aktas {\sl et al.} (H1 Collaboration), Eur. Phys. J. C {\bf 41}, 453 (2005).
\bibitem{10} M.~Cacciari and P.~Nason, Phys. Rev. Lett. {\bf 89}, 122003 (2002);\\
  M.~Cacciari, S.~Frixione, M.L.~Mangano, P.~Nason, and G.~Ridolfi, JHEP {\bf 0407}, 033 (2004).
\bibitem{11} S.~Catani, M.~Ciafoloni and F.~Hautmann, Nucl. Phys. B {\bf 366}, 135 (1991).
\bibitem{12} J.C.~Collins and R.K.~Ellis, Nucl. Phys. B {\bf 360}, 3 (1991).
\bibitem{13} L.V.~Gribov, E.M.~Levin, and M.G.~Ryskin, Phys. Rep. {\bf 100}, 1 (1983).
\bibitem{14} E.M.~Levin, M.G.~Ryskin, Yu.M.~Shabelsky and A.G.~Shuvaev, Sov. J. Nucl. Phys. {\bf 53}, 657 (1991).
\bibitem{15} S.P.~Baranov and N.P.~Zotov, Phys. Lett. B {\bf 491}, 111 (2000).
\bibitem{16} A.V.~Lipatov, V.A.~Saleev, and N.P.~Zotov, Mod. Phys. Lett. A {\bf 15}, 1727 (2000).
\bibitem{17} H.~Jung and G.~Salam, Eur. Phys. J. C {\bf 19}, 351 (2001).
\bibitem{18} S.P.~Baranov, H.~Jung, L.~J\"onsson, S.~Padhi, and N.P.~Zotov, Eur. Phys. J. C {\bf 24}, 425 (2002).
\bibitem{19} L.~Motyka and N.~Timneanu, Eur. Phys. J. {\bf C27}, 73 (2003).
\bibitem{20} A.V.~Lipatov and N.P.~Zotov, DESY 05-252.
\bibitem{21} A.V.~Lipatov and N.P.~Zotov, hep-ph/0601240.
\bibitem{22} M.G.~Ryskin and Yu.M.~Shabelsky, Z. Phys. C {\bf 61}, 517 (1994);\\
  M.G.~Ryskin, Yu.M.~Shabelsky and A.G.~Shuvaev, Z. Phys. C {\bf 69}, 269 (1996).
\bibitem{23} S.P.~Baranov and M.~Smizanska, Phys. Rev. D {\bf 62}, 014012 (2000).
\bibitem{24} Ph.~H\"agler, R.~Kirschner, A.~Sch\"afer, L.~Szymanowski and O.V.~Teryaev, Phys. Rev. D {\bf 62}, 071502 (2000).
\bibitem{25} H.~Jung, Phys. Rev. D {\bf 65}, 034015 (2002).
\bibitem{26} A.V.~Lipatov, N.P.~Zotov, and V.A.~Saleev, Yad. Fiz. {\bf 66}, 786 (2003);\\
  S.P.~Baranov, N.P.~Zotov and A.V.~Lipatov, Phys. Atom. Nucl. {\bf 67}, 834 (2004).
\bibitem{27} A.V.~Lipatov, L.~L\"onnblad, and N.P.~Zotov, JHEP {\bf 01}, 010 (2004).
\bibitem{28} H.~Jung, Mod. Phys. Lett. A {\bf 19}, 1 (2004).
\bibitem{29} M.~Hansson, H.~Jung, and L.~J\"onsson, hep-ph/0402019.
\bibitem{30} A.V.~Lipatov and N.P.~Zotov, Eur. Phys. J. C {\bf 41}, 163 (2005);\\
  A.V.~Lipatov, to be published in Yad. Fiz. (2006).
\bibitem{31} E.A.~Kuraev, L.N.~Lipatov, and V.S.~Fadin, Sov. Phys. JETP {\bf 44}, 443 (1976);\\
  E.A.~Kuraev, L.N.~Lipatov, and V.S.~Fadin, Sov. Phys. JETP {\bf 45}, 199 (1977);\\
  I.I.~Balitsky and L.N.~Lipatov, Sov. J. Nucl. Phys. {\bf 28}, 822 (1978).
\bibitem{32} M.~Ciafaloni, Nucl. Phys. B {\bf 296}, 49 (1988);\\
  S.~Catani, F.~Fiorani, and G.~Marchesini, Phys. Lett. B {\bf 234}, 339 (1990);\\
  S.~Catani, F.~Fiorani, and G.~Marchesini, Nucl. Phys. B {\bf 336}, 18 (1990);\\
  G.~Marchesini, Nucl. Phys. B {\bf 445}, 49 (1995).
\bibitem{33} V.N.~Gribov and L.N.~Lipatov, Yad. Fiz. {\bf 15}, 781 (1972);\\
  L.N.~Lipatov, Sov. J. Nucl. Phys. {\bf 20}, 94 (1975);\\
  G.~Altarelly and G.~Parizi, Nucl. Phys. B {\bf 126}, 298 (1977);\\
  Y.L.~Dokshitzer, Sov. Phys. JETP {\bf 46}, 641 (1977).
\bibitem{34} B.~Andersson {\sl et al.} (Small-$x$ Collaboration), Eur. Phys. J. C {\bf 25}, 77 (2002).
\bibitem{35} J.~Andersen {\sl et al.} (Small-$x$ Collaboration), Eur. Phys. J. C {\bf 35}, 77 (2004).
\bibitem{36} H.~Jung, Comput. Phys. Comm. {\bf 143}, 100 (2002).
\bibitem{37} J.~Kwiecinski, A.D.~Martin and A.M.~Stasto, Phys. Rev. D {\bf 56}, 3991 (1997).
\bibitem{38} M.A.~Kimber, A.D.~Martin and M.G.~Ryskin, Phys. Rev. D {\bf 63}, 114027 (2001);\\
  G.~Watt, A.D.~Martin and M.G.~Ryskin, Eur. Phys. J. C {\bf 31}, 73 (2003).
\bibitem{39} D.~Graudentz, Phys. Rev. D {\bf 49}, 3291 (1994).
\bibitem{40} A.V.~Kotikov, A.V.~Lipatov, G.~Parente, and N.P.~Zotov, Eur. Phys. J. C {\bf 26}, 51 (2002).
\bibitem{41} G.P.~Lepage, J. Comput. Phys. {\bf 27}, 192 (1978).
\bibitem{42} J.~Kwiecinski, A.D.~Martin and A.~Sutton, Phys. Rev. D {\bf 52}, 1445 (1995).
\bibitem{43} J.~Kwiecinski, A.D.~Martin and J. Outhwaite,  Eur. Phys. J. C {\bf 9}, 611 (2001).
\bibitem{44} M.~Gl\"uck, E.~Reya and A.~Vogt, Phys. Rev. {\bf D46}, 1973 (1992);\\
  M.~Gl\"uck, E.~Reya and A.~Vogt, Z. Phys. {\bf C67}, 433 (1995). 
\bibitem{45} A.V.~Lipatov and N.P.~Zotov, Phys. Rev. D {\bf 72}, 054002 (2005).  
\bibitem{46} M.A.~Kimber, A.D.~Martin and M.G.~Ryskin, Eur. Phys. J. C {\bf 12}, 655 (2001).
\bibitem{47} A.V.~Lipatov and N.P.~Zotov, DESY 05-157 [hep-ph/0507243].
\bibitem{48} C.~Peterson, D.~Schlatter, I.~Schmitt, and P.~Zerwas, Phys. Rev. D {\bf 27}, 105 (1983).
\bibitem{49} H.~Jung, Comput. Phys. Comm. {\bf 86}, 147 (1995).
\bibitem{50} B.W.~Harris and J.~Smith, Nucl. Phys. B {\bf 452}, 109 (1995).
\bibitem{51} A.V.~Kotikov, A.V.~Lipatov, and N.P.~Zotov, Eur. Phys. J. C {\bf 27}, 219 (2003).
\bibitem{52} A.V.~Kotikov, A.V.~Lipatov, and N.P.~Zotov, JETP {\bf 101}, 811 (2005).

\end{thebibliography}
\end{document}